\newif\ifANONYMOUS
\ANONYMOUSfalse

\newif\ifARXIV
\ARXIVtrue
\ARXIVfalse

\ifANONYMOUS
\documentclass[10pt,conference,anon,review]{IEEEtran}
\else
\documentclass[10pt,conference,review]{IEEEtran}
\fi

\usepackage[dvipsnames]{xcolor}
\usepackage{epsfig, endnotes, amssymb, tabularx}
\usepackage{listings, array}
\usepackage[frozencache,cachedir=mintedoutput]{minted}
\usepackage{graphicx}
\usepackage{dirtree}
\usepackage[compress]{cite}
\usepackage{caption} 
\usepackage{booktabs}
\usepackage{orcidlink}

\newcommand{\toolname}{ZTD$_{\text{JAVA}}$\xspace}
\newcommand{\tooldesign}{ZTD$_{\text{SYS}}$\xspace}
\newcommand{\maven}{Maven\xspace}
\newcommand{\mavencentral}{Maven Central\xspace}
\newcommand{\zta}{Zero-Trust Architecture\xspace}
\newcommand{\ztd}{Zero-Trust Dependencies\xspace}

\newcolumntype{C}[1]{>{\centering\arraybackslash}p{#1}}

\newif\ifDEBUG
\DEBUGtrue
\DEBUGfalse

\newif\ifKEEPKYLE
\KEEPKYLEfalse

\usepackage{booktabs} 
\usepackage{amsmath}
\usepackage{algorithm}
\usepackage{algpseudocode}
\usepackage[normalem]{ulem} 
\usepackage{xspace}
\usepackage{multirow} 
\usepackage{balance} 
\usepackage{fancyvrb} 
\usepackage{caption}

\usepackage{enumitem}
\setlist[itemize]{leftmargin=*,noitemsep,topsep=0pt}
\setlist[enumerate]{leftmargin=*}

\usepackage{etoolbox}
\makeatletter
\patchcmd{\@makecaption}
	{\scshape}
	{}
	{}
	{}
\makeatletter
\patchcmd{\@makecaption}
	{\\}
	{.\ }
	{}
	{}
\makeatother

\newcommand{\code}[1]{\texttt{{\small #1}}}

\newcommand{\ie}{\textit{i.e.,}\xspace}
\newcommand{\eg}{\textit{e.g.,}\xspace}
\newcommand{\etal}{\textit{et al.}\xspace}

\usepackage{mdframed}
\mdfsetup{skipabove=0.5\topskip,skipbelow=0.5\topskip,align=center}

\usepackage{amsthm}
\newtheorem{thm}{Theorem}

\DeclareMathSymbol{\mlq}{\mathord}{operators}{``}
\DeclareMathSymbol{\mrq}{\mathord}{operators}{`'}

\newif\ifSAVESPACE
\SAVESPACEfalse

\ifSAVESPACE

\else

\fi

\usepackage{todonotes}
\usepackage{soul}
\usepackage{fontawesome}
\usepackage{ragged2e}

\ifDEBUG
    \newcommand{\AH}[1]{\todo[color=cyan,inline]{AH:#1}}
    \newcommand{\AM}[1]{\todo[color=red,inline]{Machiry:#1}}
    \newcommand{\JD}[1]{\todo[color=yellow,inline]{JD:#1}}
    \newcommand{\SA}[1]{\todo[color=green,inline]{SA:#1}}
    \newcommand{\PA}[1]{\todo[color=orange,inline]{PA:#1}}
    
    \newcommand{\KR}[1]{\todo[color=yellow,inline]{Kyle:#1}}
    \newcommand{\LS}[1]{\todo[color=green,inline]{LS:#1}}
    \newcommand{\HP}[1]{\todo[color=green,inline]{HP:#1}}
    
    \newcommand{\TODO}[1]{\hl{#1}}
\else
    \newcommand{\AH}[1]{}
    \newcommand{\AM}[1]{}
    \newcommand{\JD}[1]{}
    \newcommand{\SA}[1]{}
    \newcommand{\PA}[1]{}
    \newcommand{\KR}[1]{}
    \newcommand{\LS}[1]{}
    \newcommand{\HP}[1]{}
    
    \newcommand{\TODO}[1]{#1}
\fi

\newcommand{\summary}[1]{
\begin{tcolorbox} [width=1.0\linewidth, colback=pink, top=1pt, bottom=1pt, left=2pt, right=2pt]
#1
\end{tcolorbox}
}

\newcommand{\definition}[1]{
\begin{tcolorbox} [width=1.0\linewidth, colback=blue!07!white, top=1pt, bottom=1pt, left=2pt, right=2pt]
#1
\end{tcolorbox}
}

\usepackage{hyperref}

\usepackage{cleveref}
\crefformat{section}{\S#2#1#3}
\crefname{figure}{Figure}{Figures}
\crefname{table}{Table}{Tables}
\crefname{theorem}{Theorem}{Theorems}
\crefname{thm}{Theorem}{Theorems}
\crefname{lemma}{Lemma}{Lemmata}
\crefname{equation}{Eqt.}{Eqts.}
\crefformat{Grammar}{Grammar #1}
\crefname{appendix}{Appendix}{Appendices}
\crefname{listing}{Listing}{Listings}

\newcommand{\myparagraph}[1]{\paragraph{#1}}
\renewcommand{\myparagraph}[1]{\vspace{0.25em} \noindent \underline{\textit{#1:}}}

\usepackage{url}

\usepackage{tcolorbox}

\makeatletter
\newcommand{\linebreakand}{%
  \end{@IEEEauthorhalign}
  \hfill\mbox{}\par
  \mbox{}\hfill\begin{@IEEEauthorhalign}
}
\makeatother

\usepackage{pifont}

 \usepackage{pifont}
 \newcommand{\cmark}{{\color{ForestGreen}\ding{51}}}%
 \newcommand{\xmark}{{\color{Maroon}\ding{55}}}%
 \newcommand{\dingone}{{\color{blue}\ding{192}}}%
 \newcommand{\dingtwo}{{\color{blue}\ding{193}}}%
 \newcommand{\dingthree}{{\color{blue}\ding{194}}}%
 \newcommand{\dingfour}{{\color{blue}\ding{195}}}%
 \newcommand{\dingfive}{{\color{blue}\ding{196}}}%

\newcommand{\vulnAnalyzed}{539\xspace}
\newcommand{\vulnManuallyReviewed}{118\xspace}

\newcommand{\packagesAnalyzed}{4387\xspace}

\newcommand{\totalMavenVulns}{4462\xspace}

\newcommand{\totalLoc}{2,284\xspace}

\newcommand{\dacapoAppCount}{22\xspace}
\newcommand{\dacapoBuiltAppCount}{18\xspace}

\begin{document}

\date{}

\title{\Large \bf Mitigating Java Supply Chain Exploits with a Runtime Permission Model}
\title{\Large \bf \toolname: Preventing Java Supply Chain RCE Exploits via a Usable Package-level Permission Manager}
\title{{\Large \bf \toolname: Preventing the next Log4Shell in Java with a Supply-chain Aware Permission Manager}}
\title{{\Large \bf Preventing Supply Chain Vulnerabilities in Java with a Fine-Grained Permission Manager}}
\title{{\Large \bf A Supply-Chain-Aware Sandbox Design for Supply Chain Vulnerability Defense}}
\title{{\Large \bf Enabling Component-level Sandboxing for Supply Chain Vulnerability Defense}}
\title{{\Large \bf A Component-level Sandbox Design for Runtime Supply Chain Vulnerability Defense}}
\title{{\Large \bf Enabling a \zta for Mitigating Software Supply Chain Vulnerability Exploitation}}

\title{{\Large \bf Enabling a \zta for Mitigating Software Supply Chain Vulnerability Exploitation in Java}}
\title{{\Large \bf \toolname: Enabling a \zta for Mitigating Software Supply Chain Vulnerability Exploitation in Java}}
\title{{\Large \bf \toolname: Designing a \zta for Mitigating Software Supply Chain Vulnerability Exploitation in Java}}
\title{{\Large \bf \toolname: A Zero Trust Design for Mitigating Software Supply Chain Vulnerabilities}}
\title{{\Large \bf \tooldesign: Design and Implementation of a \zta for Mitigating Software Supply Chain Vulnerability Exploitation in Java}}

\title{{\Large \bf Preventing Exploitation of Software Supply Chain Vulnerabilities using \ztd}}
\title{{\Large \bf {\normalsize ZTD$_{\text{JAVA}}$}: Mitigating Software Supply Chain Vulnerabilities via \ztd}}

\ifANONYMOUS
\author{
{\rm Anonymous author(s)}
}
\else



\author{
\IEEEauthorblockN{Paschal C. Amusuo\orcidlink{0000-0003-1001-525X}\textsuperscript{\dag}}
\IEEEauthorblockA{Purdue University}

\and

\IEEEauthorblockN{Kyle A. Robinson\orcidlink{0009-0004-6365-6645}}
\IEEEauthorblockA{Purdue University}

\and
    
\IEEEauthorblockN{Tanmay Singla\orcidlink{0009-0002-9108-1514}}
\IEEEauthorblockA{Purdue University}

\and
      
\IEEEauthorblockN{Huiyun Peng\orcidlink{0009-0003-2574-5316}}
\IEEEauthorblockA{Mount Holyoke College}

\linebreakand
    
\IEEEauthorblockN{Aravind Machiry\orcidlink{0000-0001-5124-6818}}
\IEEEauthorblockA{Purdue University}

\and
    
\IEEEauthorblockN{Santiago Torres-Arias\orcidlink{0000-0002-9283-3557}}
\IEEEauthorblockA{Purdue University}

\and
    
\IEEEauthorblockN{Laurent Simon\orcidlink{0000-0001-7893-547X}}
\IEEEauthorblockA{Google}

\and

\IEEEauthorblockN{James C. Davis\orcidlink{0000-0003-2495-686X}}
\IEEEauthorblockA{Purdue University}

}

\fi

\maketitle

\renewcommand{\thefootnote}{\dag}
\footnotetext{Some work performed as a Student Researcher at Google.}

\thispagestyle{empty}

\begin{abstract}


Third-party libraries like Log4j accelerate software application development but introduce substantial risk.
Vulnerabilities in these libraries have led to Software Supply Chain (SSC) attacks that compromised resources within the host system.
These attacks benefit from current application permissions approaches: third-party libraries are implicitly trusted in the application runtime.
An application runtime designed with \zta (ZTA) principles --- secure access to resources, continuous monitoring, and least-privilege enforcement --- could mitigate SSC attacks, as it would give zero implicit trust to these libraries.
However, no individual security defense incorporates these principles at a low runtime cost.

This paper proposes \emph{\ztd} to mitigate SSC vulnerabilities: we apply the NIST ZTA to software applications.
First, we assess the expected effectiveness and configuration cost of \ztd using a study of third-party software libraries and their vulnerabilities.
Then, we present a system design, \tooldesign, that enables the application of \ztd to software applications and a prototype, \toolname, for Java applications. 
Finally, with evaluations on recreated vulnerabilities and realistic applications, we show that \toolname can defend against prevalent vulnerability classes, introduces negligible cost, and is easy to configure and use.

\end{abstract}

\section{Introduction}

\begin{figure}[t]
 \centering
\captionsetup{font=small}
 \includegraphics[width=0.95\columnwidth]{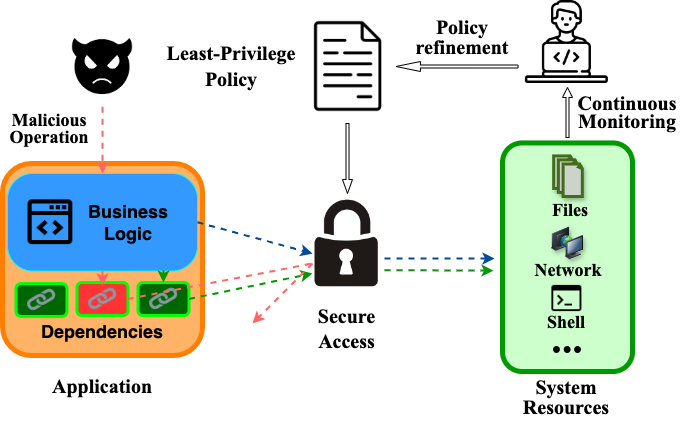}
 \captionof{figure}{
   The \textit{Zero-Trust Dependencies} (ZTD) concept.
   To mitigate attacks exploiting vulnerable dependencies, a ZTD system provides \textit{secure access} via runtime authorization, makes authorization decisions using a \textit{least-privileges access} policy, and facilitates \textit{continuous monitoring} of unexpected accesses.
 }
 \label{fig:ztd-overview}
\end{figure}

Integrating third-party libraries (\ul{TPLs}) such as Log4j as dependencies accelerates software application development but introduces risks~\cite{thompson_reflections_1984, nikiforakis_you_2012, zimmermann_small_2019}.
Dependencies are implicitly trusted by default and execute with the application's permissions.
Vulnerabilities in dependencies, termed \textit{software supply chain (\ul{SSC}) vulnerabilities}~\cite{yan_estimating_2021}, may cause undesirable application behavior~\cite{ali_armor_2020}.
They result in SSC attacks~\cite{noauthor_equifax_nodate_2}.
To mitigate analogous risks from assets in cloud systems and corporate networks,
  the US National Institute of Standards and Technology (NIST) has recommended~\cite{rose_zero_2020_2} (and industry~\cite{google_zero_trust, microsoft_zero_trust, ibm_zero_trust} and academia have adopted~\cite{dsilva_building_2021, bertino_zero_2021, phiayura_comprehensive_2023, fernandez_critical_2024})
  the \textbf{\zta} (\ul{ZTA}) to place zero implicit trust in system actors.
We propose applying ZTA \textit{within} a software application to mitigate the risk of SSC vulnerabilities (\cref{fig:ztd-overview}).

Many security defenses reduce the risks of using TPLs within an application.
However, none sufficiently mitigates SSC vulnerabilities as they do not enable the ZTA principles of secure resource access, continuous monitoring, and least privileges for the application's dependencies and at low runtime cost (\cref{sec:RelatedWork}).
\emph{Application-level} sandboxes~\cite{noauthor_selinux_nodate_2, noauthor_apparmor_nodate_2, noauthor_docker_2022_2, noauthor_security_nodate_2} do not operate on dependencies.
\emph{Import-restriction-based}~\cite{de_groef_nodesentry_2014, ferreira_containing_2021, vasilakis_preventing_2021} and \emph{debloating-based}~\cite{vasilakis_supply-chain_2021, pashakhanloo_pacjam_2022} approaches do not validate when dependencies access resources.
Existing \emph{isolation-based} techniques~\cite{sun_nativeguard_2014, abbadini_cage4deno_2023, vasilakis_breakapp_2018, seo_flexdroid_2016}
do not enable discovery of dependencies' least privileges, and they introduce high runtime costs.

This paper proposes \emph{\ztd} (\cref{fig:ztd-overview}), a concept that applies NIST's \zta to an application's dependencies to prevent software supply chain attacks. 
We begin with a feasibility study of the \zta idea, clarifying its suitability and design considerations through a study of software supply chain vulnerabilities and application dependency chains (\cref{sec:understanding-sscv}).
Motivated by our findings, we state the threat model (\cref{subsec:threat-model}), propose \ztd as a mitigation (\cref{subsec:design-reqs}), and design a ZTD system, \tooldesign to
  enable secure resource access (\cref{subsec:secure-resource-access}),
  discover least privileges (\cref{subsec:discovery-least-privileges}),
  and
  continuously monitor dependency behaviors (\cref{subsec:continous-monitoring}).

We implement a prototype, \toolname, for Java applications (\cref{sec:ztd-java-implementation}).
We evaluate
  its effectiveness on recreated vulnerabilities,
  micro and macro performance costs,
  and
  the effort to audit policies of real applications. 
Our results show that fine-grained enforcement of a library's observed behavior (representing its least privileges) prevents all reproduced vulnerability exploits.
\toolname introduces much lower overhead than the state-of-art for Java.
Additionally, only an average of five dependencies in each application require explicit policy specification.

\textit{In summary, we contribute}:

\begin{itemize}
\item A feasibility study of applying \zta principles within a software application (\cref{sec:understanding-sscv}). 
\item The \textit{\ztd} concept and our \tooldesign design for mitigating SSC vulnerabilities (\cref{sec:design-implementation}). 
\item A prototype for Java, \toolname (\cref{sec:ztd-java-implementation}), and an evaluation of its cost and effectiveness on realistic Java applications (\cref{sec:evaluation}). 
\end{itemize}

\myparagraph{Significance}
Operationalizing a recommended security architecture will enable its adoption in software engineering, leading to more secure applications.
Our feasibility study, design, and evaluation will inform engineers of the utility and cost of adopting this security architecture in their applications.

\section{Background} \label{sec:Background}

Here we 
  define software supply chains and vulnerabilities (\cref{subsec:ssc-vulns}),
  introduce \zta as a mitigation (\cref{subsec:zero-trust-arch}),
  and
  discuss limitations of existing defenses (\cref{sec:RelatedWork}).

\subsection{Software Supply Chains (SSC) and Vulnerabilities}
\label{subsec:ssc-vulns}

From a technical standpoint, \textit{Software Supply Chains} encompass the systems and devices involved in producing a final software product~\cite{okafor_sok_2022}. This includes the application's source code, build tools, final packaged artifact, and third-party libraries used as dependencies~\cite{Supply-chain-Levels-for-Software-Artifacts}.
The use of third-party libraries introduces new attack surfaces. They can be deliberately~\cite{Naraine_2022, Wallen_2021, SolarWinds-Hack_2021} or unintentionally (\eg \cref{listing:sscv-snippet}) introduced. These vulnerabilities are \textit{Software Supply Chain (SSC) vulnerabilities}~\cite{yan_estimating_2021} and they enable \textit{SSC Attacks}~\cite{noauthor_equifax_nodate_2, solarwinds_setting_2021} that compromise the host system and lead to consequences such as execution of arbitrary code, file manipulation and exfiltration of sensitive data.
SSC attacks typically succeed because vulnerable dependencies inherit the application's permissions, obtaining more privileges than they need in typical use~\cite{ferreira_containing_2021}. 

SSC vulnerabilities can be mitigated by securing the structure and use of a software supply chain.
Per Okafor~\etal~\cite{okafor_sok_2022},
  secure supply chains need
  transparency (where dependencies come from and what risks they contain),
  validation (reliability of dependencies),
  and
  separation (dependency isolation to avoid domino effects).
\textit{Transparency} can be improved through techniques such as software bills of materials (SBOMs)~\cite{noauthor_software_nodate_2} and the Open-Source Vulnerabilities project~\cite{noauthor_osv_nodate_2}.
\textit{Validity} is promoted through software signing~\cite{newman_sigstore_2022, torres-arias_-toto_2019} to verify the provenance of third-party libraries and the end-to-end security model~\cite{hammi_software_2023} that prevents unauthorized alteration of source code.
Our zero-trust dependencies concept, inspired by NIST's zero-trust architecture, improves \textit{separation} between dependencies.

\begin{listing}
  \centering
  \captionsetup{font=small}
  \caption{
   An exploit for a code Injection vulnerability (CVE-2022-22963) in Spring Cloud Function (SCF)~\cite{spring_cloud_function}.
   On lines 8-11, an attacker executes a malicious command by using SCF's routing-expression functionality.
   Executing shell commands is not a privilege SCF needs, and thus would be blocked by a \zta.
  }
  \label{listing:sscv-snippet}
  \begin{tcolorbox} [width=\linewidth, colback=white!30!white, top=1pt, 
  bottom=1pt, left=2pt, right=2pt]
\begin{minted}[
    fontsize=\scriptsize,
    linenos,
    gobble=2, % Remove unnecessary indentation -- the line numbers make this clear enough
    xleftmargin=0.5cm, % Otherwise we start in the left margin...
    escapeinside=||
]{c}

public void exploitTarget() {
  String command = // exploit command

  HttpUrlConnection con = 
    new URL(URL).openConnection();

  con.setRequestMethod("POST");
  con.setRequestProperty(
  "spring.cloud.function.routing-expression", 
  "T(java.lang.Runtime).getRuntime().exec
  (\"+command+\")");

  int response = con.getResponseCode();
}

  \end{minted}
\end{tcolorbox}
\end{listing}


\subsection{\zta as a Conceptual Framework}
\label{subsec:zero-trust-arch}

\cref{subsec:ssc-vulns} suggests that SSC vulnerability exploits succeed because dependencies are granted permission to access more resources than they typically need.
The \zta (ZTA)~\cite{rose_zero_2020_2, kindervag_build_2010} protects access to resources in the context of a network and may analogously effectively mitigate SSC vulnerabilities.
As defined by the USA's National Institute of Standards and Technology (NIST) in SP-800-207~\cite{rose_zero_2020_2}, \zta
  is a set of cybersecurity concepts focused on protecting
  services (\textit{resources}) within a system,
  grants no implicit trust to any user or assets (\textit{subjects})
  and requires subjects to gain explicit authorization to access any resource. 
Following Google~\cite{google_zero_trust}, we summarize ZTA in 3 principles:

\begin{enumerate}
\item \textit{Secure and Context-based Access:} Every access to a resource should be authorized. 
\item \textit{Continous Monitoring:} Organizations should monitor the state and activities of the subjects and use the insights gained to improve the creation and enforcement of policies.
\item \textit{Least-Privilege Policy Enforcement:} 
Access policies should grant minimum access rights to subjects. 
\end{enumerate}

The \zta has been applied to security-sensitive domains~\cite{syed_zero_2022} including network infrastructure~\cite{ramezanpour_intelligent_2022, federici_zero-trust_2023} and cloud systems~\cite{google_zero_trust, microsoft_zero_trust, dsilva_building_2021}.
When applied to software applications, ZTA can mitigate SSC vulnerability exploits as it authorizes only legitimate access to operating system resources, based on a dependency's minimum privileges set.

\subsection{Limitations of Existing Application Security Defenses}
\label{sec:RelatedWork}

Existing security defenses reduce some risks from third-party libraries, but do not prevent SSC vulnerability exploits as they do not enforce zero-trust principles on all dependencies.
We summarize these techniques in~\cref{tab:related-works-sandboxes}.

\begin{table}
    \centering
    \captionsetup{font=small}
    \caption{
    Analysis of existing security defenses by ZTA principles.
    Columns indicate if each technique family 
      provides secure resource access,
      supports least priv. discovery and enforcement,
      enables continuous monitoring for dependencies,
      and
      has low runtime costs.
    }
    \label{tab:related-works-sandboxes}
    \begin{tabular}{C{2.5cm}C{1.1cm}|C{0.74cm}C{0.64cm}C{0.87cm}|C{0.5cm}}
    \toprule
         \textbf{Family} & \textbf{Design} & \textbf{Res. Access} & \textbf{Lst-Prv. Enf.} & \textbf{Dep. Mon.} &  \textbf{Low Cost} \\
     \toprule

     \cite{noauthor_selinux_nodate_2, noauthor_seccomp_nodate, noauthor_security_nodate_2, noauthor_apparmor_nodate_2} & Access control & \cmark & \xmark & \xmark & \cmark  \\
     \cite{sultana_towards_2022, bittau_wedge_2008, wu_automatically_2013} & App decomp. & \cmark & \xmark & \xmark & \cmark \\
     \cite{vasilakis_preventing_2021, ferreira_containing_2021, ohm_you_2023} & Import restrict. & \xmark & \xmark & \cmark & \cmark  \\
     \cite{vasilakis_supply-chain_2021, pashakhanloo_pacjam_2022} & Debloating & \xmark & \xmark & \cmark  & \cmark   \\
     \cite{sun_nativeguard_2014, abbadini_cage4deno_2023, vasilakis_breakapp_2018, seo_flexdroid_2016} & Isolation & \cmark & \xmark & \cmark & \xmark \\
     \midrule
     \toolname & Access control & \cmark & \cmark & \cmark & \cmark  \\

     \bottomrule    
    \end{tabular}
\end{table}

\emph{Application-level} defenses enforce security policies on the application~\cite{noauthor_selinux_nodate_2, noauthor_seccomp_nodate, noauthor_security_nodate_2} or its logical units~\cite{sultana_towards_2022, bittau_wedge_2008, wu_automatically_2013}. 
However, as dependencies inherit the application's permissions, vulnerable dependencies may compromise sensitive system resources by leveraging permissions provided to the application. In addition, specifying security policies that enforce least privileges in modern applications is difficult and error prone~\cite{schreuders_state_2013, maass_theory_2016_2}. The Java Security Manager (JSM)~\cite{noauthor_security_nodate_2}, an application-level defense for Java applications, allows specifying fine-grained policies for secure resource access in Java applications. It was deprecated in 2021 due to lack of use, ascribed to its brittle permission model, difficult programming model, and poor performance~\cite{sean_mullan_jep_nodate_2}.

\emph{Dependency-level} defenses operate on dependencies and can constrain their behaviors at runtime. 
We distinguish three kinds.
\emph{Import-restriction-based} defenses \cite{vasilakis_preventing_2021, ferreira_containing_2021, ohm_you_2023} intercept dependencies as they are imported into an application and remove unauthorized functions. 
Meanwhile, \emph{debloating-based} defenses~\cite{vasilakis_supply-chain_2021, pashakhanloo_pacjam_2022} remove unused or vulnerable functions from a library's source code. 
These designs prevent malicious code introduction in dependencies and introduce lower run-time costs. However, they do not validate dependency resource accesses and cannot prevent exploits of accidental vulnerabilities where the attacker only controls data in the app.
\textit{Isolation-based} defenses~\cite{vasilakis_breakapp_2018, seo_flexdroid_2016, sun_nativeguard_2014, abbadini_cage4deno_2023} execute dependencies in isolated compartments and can enable secure access to resources. However, they face low adoption rates~\cite{ladisa_sok_2023_2}, in part, due to their high performance overhead and the effort required to discover the least privileges of a compartment.

\section{\zta Feasibility Analysis}
\label{sec:understanding-sscv}

Previous works have demonstrated
  the feasibility of secure context-based access~\cite{vasilakis_breakapp_2018, seo_flexdroid_2016} (ZTA principle 1) and
  dependency monitoring~\cite{seo_flexdroid_2016, ferreira_containing_2021} (ZTA principle 2). However, no relevant security defense (\cref{sec:RelatedWork}) has assessed the feasibility of least-privilege definitions for dependencies.

\JD{Reminder, this introductory section now explains why we need to do the feasibility analysis. Several reviewer made individual comments about this I think, so mention this part in the letter.}
This section measures two distinct aspects of applying the principle of least-privilege policy on dependencies. First, ZTA requires that SSC vulnerability exploits on a dependency involve \textit{resources not needed by that dependency}, so that a least-privilege policy would mitigate these vulnerabilities without impacting the application. 
Second, ZTA requires that \textit{the cost of correctly configuring least-privilege policies be reasonable}, \eg that only a few dependencies need privileges and require policy specification.
Thus we ask:

\begin{itemize} [leftmargin=25pt, rightmargin=5pt]
    \item [\textbf{RQ1}] What proportion of \textit{Java SSC vulnerabilities} could be mitigated by enforcing least-privilege policies?
    \item [\textbf{RQ2}] What proportion of \textit{dependencies} will need explicit policy authorization in an application?
\end{itemize}

To answer these questions, we measure SSC vulnerabilities and third-party libraries in the Java ecosystem.

\textbf{Why Java?}
We situate this feasibility study, and our subsequent embodiment of ZTD, within Java.
Java is a popular programming language~\cite{noauthor_github_nodate_2, noauthor_tiobe_nodate_2}, 
and SSC attacks have been the most impactful in the Java ecosystem. 
For example, the Log4j vulnerability affected millions of Java applications~\cite{CISALog4jReport2022} with estimated costs in the billions~\cite{Log4ShellSCMag2022}.
While there have been SSC attacks in other ecosystems like JavaScript and Python~\cite{CycodeESLint2022}, their impact has been smaller due to their smaller server-side industry footprints.

\textbf{Novelty}:
Prior works have studied the life cycle~\cite{Alfadel2021, zhang2023mitigatingpersistenceopensourcevulnerabilities} and dependency-tree propagation behaviors~\cite{kaplan2021, Mir2023} of SSC vulnerabilities.
This study instead describes how SSC vulnerabilities access resources in third-party libraries.

\subsection{Methodology}
\label{sscv-study-datasets}


\subsubsection{RQ1}


We evaluated RQ1 by assessing the proportion of Java SSC vulnerabilities that involved access to operating system resources and examining whether they can be mitigated by coarse or fine-grained least-privilege policy enforcement.

We study recent and high-severity vulnerabilities in popular \mavencentral libraries, as they represent prevalent and high-impact threats to applications.
We obtained the \totalMavenVulns vulnerabilities from the Open-Source Vulnerabilities (OSV) database~\cite{noauthor_osv_nodate_2}.
We filtered for vulnerabilities that were published within the last 5 years, had a CVSSv3 rating of high or critical, and affected the top 10,000 depended-upon \mavencentral libraries from the Ecosystems database~\cite{ecosystems_db}. 
This yielded \vulnAnalyzed vulnerabilities in 252 unique libraries.

To answer RQ1, we randomly analyzed \vulnManuallyReviewed of the \vulnAnalyzed vulnerabilities (22\%)~\footnote{This represents an 8\% confidence interval at a 95\% confidence level.}.
This number is comparable to the number of vulnerabilities prior works~\cite{amusuo_systematically_2023, jin_understanding_2012} studied.
First, we classified them using
  the taxonomy of web security vulnerabilities~\cite{al-kahla_taxonomy_2021} and
  their exploitation impact~\cite{ohm_backstabbers_2020}.
We report the proportion of vulnerabilities whose impact enables malicious access to resources.
For these vulnerabilities, we 
 identified the resources they expose and 
 the API that the vulnerable code uses to access the resource.
We also compiled a list of APIs that vulnerable libraries used to access each resource type.
Next, we used CodeQL~\cite{noauthor_codeql_nodate_2} and the compiled list of APIs to search for API calls that access the exposed resource in the affected library.
If no calls were found, a coarse least-privilege policy that denies access to the specific resource type would prevent an exploit.
However, if API calls were found, a fine-grained policy would be needed that only authorizes access to legitimately needed resource objects.


\subsubsection{RQ2}
RQ2 estimates the proportion of dependencies that access sensitive resources and would need policy authorization.
From a library perspective, we measured the proportion of popular \mavencentral libraries with direct access to OS resources.
From an application perspective, we measured the proportion of dependencies in real applications that access OS resources during runtime.
The results provide an upper bound and a common estimate of the number of dependencies that would require explicit policies.
High values would threaten the feasibility of ZTA when applied to software dependencies.

For the upper bound, we obtained 23,569 \mavencentral libraries sorted by their number of dependents, from the Ecosystems~\cite{ecosystems_db} database for our analysis. 388 libraries could not be cloned, and 18,495 failed to build due to missing dependencies.  Similar to RQ1, we used CodeQL~\cite{noauthor_codeql_nodate_2} to analyze the remaining 4,686 libraries, identify sensitive API calls, and report the percentage of libraries that access each type of resource. 
For the common estimate, we used the 2023 DaCapo benchmark suite for Java, consisting of 22 applications, including a web server (Tomcat), web application framework (Spring), IDE (Eclipse), database (Cassandra), and message bus (Kafka), and workloads that imitate industry needs~\cite{blackburn_dacapo_2006}. Four applications did not build or run and we could not get the list of dependencies for nine applications. We analyzed the dependencies of the remaining nine applications using our CodeQL scheme. We instrumented the nine applications to record the resources each dependency accessed, executed the applications, and compared the resources that the dependencies accessed at runtime with the resources that the dependencies can access as indicated in their CodeQL results.

\subsection{Results}
\label{subsec:sscv-results}

\subsubsection{RQ1: Will least-priv policies mitigate SSC vulns?}

\summary{\textbf{Finding 1 (RQ1)}: 46\% of Java SSC vulnerabilities can compromise an operating system resource. 58\% (of 46\%) can be mitigated using a coarse-grained least-privilege policy. 42\% require finer-grained least-privilege policies that can control the resource objects they access.}

\cref{tab:vulnerability-count} shows the different vulnerability classes in Java third-party libraries and the operating system resources they expose.
Five vulnerability classes, comprising 46\% of vulnerabilities, provide access to the file system, network, and shell. They enable a malicious actor to manipulate files, exfiltrate data, and execute commands remotely. 
From a study of malicious libraries, Ohm~\etal~\cite{ohm_backstabbers_2020} found that these objectives represented 97\% of the analyzed attackers' goals.

\begin{table}
    \centering
    \captionsetup{font=small}
    \caption{
    Vulnerability classes in Java third-party libraries. The table shows the consequences of exploitation, the required access to the operating system, and the count. We also cite Common Weakness Enumerations (CWEs) associated with each vulnerability class.
    }
    \begin{tabular}{p{2.2cm}C{2.5cm}C{1cm}C{0.3cm}C{0.8cm}}
    \toprule
         \textbf{Vuln. Class} & \textbf{Consequence} & \textbf{OS Resource} & \textbf{\#} & \textbf{Perc} \\
    \midrule
    Deserialization~\cite{cwe_502} & Remote Code Exec. & Shell, FS & 21 & 18\% \\
    Access Control~\cite{cwe_863} & Unauth. App Access & N/A & 21 & 18\%  \\
    Resource Exhaust.~\cite{cwe_770} & Denial of Service & N/A & 18 & 15\%  \\
    Code Injection~\cite{cwe_94} & Remote Code Exec. & Shell & 13 & 11\% \\
    Path Traversal~\cite{cwe_22} & File Read/Write & FS & 12 & 10\% \\
    XSS, CSRF~\cite{cwe_79} & Unauth. Web Access & N/A & 11 & 9\% \\
    XXE~\cite{cwe_611} & Data Exfiltration & FS, Net & 5 & 4\% \\
    Command Injec.~\cite{cwe_78} & Remote Code Exec. & Shell & 4 & 3\% \\
    Impl. flaw~\cite{cwe_248} & Varies & N/A & 8 & 7\%  \\
    Others & Varies & N/A & 5 & 4\%   \\
    \midrule
    Total & - & - & 118 & 100\%  \\
 \bottomrule
    \end{tabular}
    \label{tab:vulnerability-count}
\end{table}

\cref{tab:resource-usage} (second section) shows the number of vulnerabilities that expose different resource types and the proportion of vulnerable libraries that need access.
Some vulnerabilities expose multiple resources.
In 27 of 36 cases (75\%), the shell is exposed when not needed.
A coarse-grained policy can prevent these accesses.
In contrast, in the 25 instances where file I/O or network access is exposed, the vulnerable component did not need the respective resource in only 5 instances (20\%). 
For cases where the affected library needs the exposed resource, a more fine-grained notion of privilege would be needed, \eg permitting access to some files or remote hosts but not others.
These observations guide our policy definition in~\cref{subsubsec:perm-specs}.


\begin{table}
    \centering
    \captionsetup{font=small} 
    \caption{
    Operating system resources exposed by vulnerabilities. RQ1 (Vulns) shows the vuln. count and indicates whether the vuln. library can access the resource. RQ2 (\maven) first column shows the proportion of \packagesAnalyzed analyzed \mavencentral libraries that can access these resources directly. RQ2 (\maven) second column (a/b) shows the number of dependencies (out of 103) that can access a resource (a) and that accessed the resource within a DaCapo application (b). 
    }
    \begin{tabular}{p{1.5cm}|*{3}{C{0.8cm}}|*{2}{C{0.9cm}}}
    \toprule
    & \multicolumn{3}{c|}{\textbf{RQ1 (Vulns)}} & \multicolumn{2}{c}{\textbf{RQ2 (\maven)}} \\
    \cmidrule(lr){2-4} \cmidrule(lr){5-6}
    \textbf{OS Resource} & \textbf{Vuln Count} & \textbf{No Access} & \textbf{Access} & \textbf{\% Total} & \textbf{Caps usage} \\
    \midrule
    File read & 7 & 1 & 6 & 33\% & 46/4 \\
    File write & 11 & 1 & 10 & 28\% & 46/1 \\
    Network connection & 7 & 3 & 4 & 12\% & 6/0 \\
    Shell execution & 36 & 27 & 9 & 9\% & 15/0 \\
    \bottomrule
    \end{tabular}
    \label{tab:resource-usage}
\end{table}

\subsubsection{RQ2: How many deps. need explicit least-priv. policies?}
\summary{\textbf{Finding 2 (RQ2)}: In our upper bound estimate, 33\%, 12\%, and 9\% of \mavencentral libraries would require permissions to directly access file, network, and shell resources, respectively. However, only 4\% of the 103 dependencies in 9 applications required explicit least-privilege policies.}

Our library-focused result is shown in \cref{tab:resource-usage} (last two columns). 
First, less than 35\% of the popular third-party libraries of the \mavencentral ecosystem directly access any particular resource. For example, while 36/55 vulnerabilities (65\%) provide access to the Shell, only 16\% (9/55) of the vulnerable libraries and 9\% of popular \mavencentral libraries require Shell permissions. 

From an application perspective, in the 9 studied applications (comprising 103 dependencies), while up to 46 dependencies (45\%) could access the file system, only 4 dependencies and 1 dependency read or wrote to a file within the application. Similarly, while 15 dependencies could execute shell commands, none of the dependencies executed a shell command while executed by the application.

\subsection{Discussion}

\myparagraph{ZTA Effectiveness}
Per \cref{tab:vulnerability-count}, 55/118 of SSC vulnerabilities allow resource compromise and can be mitigated by a secure access control system.
Of these, 32 (58\%) could be addressed with a coarse-grained least-privilege policy based on resource type access, while 23 (42\%) require finer-grained policies based on specific resource objects.
Recall that existing SSC security defenses support only coarse-grained privileges (\cref{sec:RelatedWork}) and thus are ineffective against 42\% of SSC vulnerabilities.

\myparagraph{ZTA Configuration Cost}
To effectively apply ZTA, one must specify policies for application dependencies (defaulting to no trust).
Hence, the configuration cost depends on the number of dependencies that require policy specification.
A high configuration cost would make ZTA impracticable in this context.
\cref{tab:resource-usage} shows that most dependencies do not require access to sensitive resources.
When they might, the access is typically not used by their dependents.

\myparagraph{Study Limitations}
We note two cases in which our CodeQL measurement will fail.
First, a resource might be accessed with an API not covered by our queries, which we mitigated by building queries based on real exploits.
Second, a resource might be accessed through indirection (\eg callbacks or Java reflection).
This threat is a consequence of using static analysis~\cite{ferreira_containing_2021, vasilakis_preventing_2021}, which we chose for its scalability.

\section{\ztd: Concept and Design}
\label{sec:design-implementation}


This section introduces \ztd (ZTD) as a security architecture that mitigates SSC vulnerabilities.
We outline the threat model in \cref{subsec:threat-model}, define \ztd in \cref{subsec:design-reqs}, and discuss design considerations for different \tooldesign components (\cref{subsec:ztd-policy-auth} -- \cref{subsec:continous-monitoring}). 


\subsection{System and Threat Model} \label{subsec:threat-model}

\noindent
\textbf{System Model:} 
An application depends on a vulnerable third-party library that can access system resources using untrusted (\ie user-controlled) application data (shown in \cref{sec:understanding-sscv}).
The resource access operation can be initiated by the vulnerable dependency but executed directly (using function calls or callbacks) or asynchronously (in independent or child threads) by other dependencies or the application itself~\cite{seo_flexdroid_2016}.

\vspace{0.1cm}
\noindent
\textbf{Threat Model:}
We include some threats and exclude others. 
\begin{itemize}
    \item \emph{In-scope:}
The attacker controls either the source code of the vulnerable dependencies or the data that the application passes to the vulnerable dependency. Hence, they can compromise the confidentiality and integrity of the host system by exploiting the vulnerabilities in the dependency.

    \item \emph{Out-of-scope:} We do not consider threats from the SSC vulnerabilities that cause denial-of-service or unauthorized application access shown in \cref{tab:vulnerability-count}. Denial-of-service attacks can compromise the application's availability but
    they can be easily detected by monitoring the application~\cite{zare_techniques_2018}. Sensitive application functions should require additional authorization and should not be accessible from code injection attacks. 
\end{itemize}

\noindent
This threat model is stronger than the import-restriction and debloating-based models (\cref{sec:RelatedWork}) as it also prevents exploits that only control data passed into the dependency.

\subsection{The \ztd Concept}
\label{subsec:design-reqs}


NIST's \zta protects resources within a network (\cref{subsec:zero-trust-arch}).
In this paper, we define \textit{resources}~\cite{rose_zero_2020_2} in the software application context as
data and computing services provided by the operating system (OS) that an application can operate on.
Data services include the file and network system, which enable access to confidential information that the OS or its applications produce or use.
Computing services include the shell execution system that executes shell commands in the OS.
Operations refer to Read, Write, and Execute (RWX) operations that can be performed on these resources.

To mitigate the use of SSC vulnerabilities to compromise system resources, we introduce \textit{\ztd} as an adaptation of the NIST \zta (\cref{subsec:zero-trust-arch}) to the context of software applications.

\definition{\textbf{\ztd} is a software engineering paradigm that grants \ul{no implicit trust} to dependencies in an application. It requires that dependencies possess explicit authorization to create or operate on resources within the application's operating system.
}

\ztd involves three principles:


\begin{enumerate}
    \item \textbf{ZTD-P$_1$: Secure Access:} The access of dependencies to resources should be authorized using the configured access policies and the application's execution context.
    
    \item \textbf{ZTD-P$_2$: Least Privileges Enforcement:} The access policies should reflect the least privilege set that dependencies require to operate within the application.

    \item \textbf{ZTD-P$_3$: Continous Monitoring:} The software engineers should be able to continuously monitor the resource access of dependencies and use the insights gained to improve threat intelligence and policy specification.
    
\end{enumerate}

\noindent
The ZTD paradigm's \textbf{security guarantee} is:
  an application's dependencies cannot compromise the \textit{Confidentiality} or \textit{Integrity}~\cite{samonas_cia_2014} of the operating system's resources. 
Consequently, ZTD reduces the risks that applications face from using third-party libraries.



Dependency-level security defenses in \cref{sec:RelatedWork} do not enforce the three ZTD principles in an application. Hence, we also introduce the \textit{ZTD System} (or \tooldesign for short).

\noindent
\definition{A \textbf{ZTD System} is the set of access control policy designs, algorithms, and tools that enable software engineers to apply the ZTD paradigm in their applications.}

\subsection{ZTD Policy Design}
\label{subsec:ztd-policy-auth}


\begin{listing}
  \centering
  \captionsetup{font=small}
  \caption{
   Policy Specification file. It supports coarse-grained (\eg `fs.read') and fine-grained (\eg `fs.read.allowed') permissions.
  }
  \label{listing:perm-specification}
  \begin{tcolorbox} [width=\linewidth, colback=white!30!white, top=1pt, 
  bottom=1pt, left=2pt, right=2pt]
\begin{minted}[
    fontsize=\scriptsize,
    linenos,
    gobble=2, % Remove unnecessary indentation -- the line numbers make this clear enough
    xleftmargin=0.5cm, % Otherwise we start in the left margin...
    escapeinside=||
]{c}
# Coarse-grained Policy
{
  "com.app.bar": {
    "fs.read": true,
    "fs.read.denied": ["/tmp","/sensitive"],
    ...
  }

# Fine-grained Policy
  {
  "com.foo.baz": {
    "fs.write": true,
    "fs.write.allowed": ["app/logs"],
    "runtime.exec": true,
    "runtime.exec.transitive": ["whoami"]
  },
  ...
}



  \end{minted}
\end{tcolorbox}
\end{listing}

\ztd require policies that specify the least privileges of dependencies. We propose: 

\subsubsection{ZTD Permission Model}
\label{subsubsec:components-perm-model}

The ZTD permission model is designed to protect resources, within the operating system, that vulnerability exploitation may expose. We use the access matrix terminology from Sandhu \etal~\cite{sandhu_access_1994} to describe the ZTD permission model.
Permissions are given to dependencies to perform operations on resource objects.
\cref{tab:ztd-resource-types} shows the three resource types supported and the operations that can be performed.
We focus on high-risk resource types to simplify policy files and reduce permission verification frequency.
These resource types are high-risk because they enable code execution, data exfiltration, and file manipulation attacks, \ie the major targets of supply chain attacks~\cite{ohm_backstabbers_2020}.

\begin{table}
    \centering
    \captionsetup{font=small}
    \caption{
    Resource types supported by the ZTD permission model.
    }
    \begin{tabular}{p{2cm}p{2.5cm}p{2cm}}
    \toprule
         \textbf{Resource Type}&  \textbf{Operation}& \textbf{Resource Objects}\\
    \toprule
         File System &  Read, write & Files\\
         Network System &  Connect & Network URLs\\
         Shell System & execution & Shell commands\\
 \bottomrule
    \end{tabular}
    \label{tab:ztd-resource-types}
\end{table}

\subsubsection{Permission Specification Granularity}
\label{subsubsec:perm-specs}

Similar to the Java Security Manager's permission design~\cite{oracle2014permissions}, \tooldesign allows specifying coarse and fine-grained permissions. \cref{listing:perm-specification} shows samples of a policy file. 

\textit{Coarse-grained Permissions} authorize a dependency to access any object of the specified resource type. For example, a coarse file read permission (\code{fs.read}) in \cref{listing:perm-specification} allows the \code{com.app.bar} dependency to read any file in the OS, except files specified in \code{fs.read.denied}.

\textit{Fine-grained Permissions} authorize a dependency to access explicitly authorized resources. \tooldesign provides \textit{direct} and \textit{transitive} permissions that specify if an object can be accessed directly or through another dependency.
For example, \cref{listing:perm-specification} grants the \code{com.foo.baz} dependency the permission to write files, but only allows writing to those in \code{fs.write.allowed}.
The dependency is also granted the transitive runtime execution permission, so it can transitively execute commands specified in \code{runtime.exec.transitive}. Hence, while \code{com.foo.baz} cannot execute the \code{whomai} command directly, it can use another dependency that has direct permission to execute the command.

\subsection{ZTD-P$_1$: Secure and Context-Sensitive Access Control}
\label{subsec:secure-resource-access}

The System Model in \cref{subsec:threat-model} shows that multiple dependencies in an application can interact to perform a given task, using direct or asynchronous calls. \tooldesign relies on a context-sensitive access control system that considers this set of interacting dependencies when making access decisions and prevents malicious dependencies from leveraging permissions granted to other dependencies.
This section and \cref{listing:check-perm-algo} shows how \tooldesign handles components that interact directly and asynchronously.
 
\subsubsection{Handling Direct Dependency Interactions}
\label{subsubsec:perm-model-props}

A dependency can call functions from another dependency to access operating system resources.
These dependencies will be contained in the executing thread's call stack.
Hence, \tooldesign grants access to a resource if all dependencies on the call stack have the necessary direct or transitive permissions to access the resource (\cref{listing:check-perm-algo} lines 4-5, 20-28). 
This approach ensures that no dependency has permission to access any resource on its own and prevents malicious dependencies from exploiting privileged dependencies to access unauthorized resources.

\subsubsection{Handling Indirect Dependency Interactions}
\label{subsubsec:thread-perm-transfer}

In multithreaded applications, a dependency can delegate a resource access operation to another dependency or the application (\textit{delegates}).
In a simpler case, the delegate may create a child thread to access the resource~\cite{seo_flexdroid_2016}.
Then, the call stack will not contain the dependency that initiated the operation.
\tooldesign handles this interthread delegation scenario by using the permissions of dependencies in the parent and child thread's call stack to authorize access (\cref{listing:check-perm-algo} lines 8-17).
When a child thread is created, its parent's dependency policies are saved. The saved policies are retrieved and verified whenever the child thread accesses any resource. 
As reported in \cref{subsec:guarantees-limitations}, this design does not handle more complex delegations involving two independent threads or processes.


\subsubsection{Enforcing ZTD Policies}
\label{subsec:enforcing-ztd}


When access to a resource is denied, \tooldesign provides two enforcement options that balance the reliability needs of an application against the risk appetite of the engineering team:
\begin{enumerate}
    \renewcommand{\labelenumi}{\alph{enumi})}   
    \item \textit{Fatal Enforcement mode:} This option raises an exception to block the unauthorized action. Software engineers working on security-sensitive applications with low-risk tolerance may prefer this mode. They can implement exception-handling routines to prevent disruption to their applications.
    \item \textit{Non-fatal Enforcement mode:} This option does not interfere with the application. Instead, \tooldesign sends an unauthorized access alert to the application maintainers, who will investigate the incident and begin any necessary remedial actions. Software engineers maintaining applications with high availability needs or in low security-sensitivity environments may prefer this mode.
\end{enumerate}


\begin{listing}
  \centering
  \captionsetup{font=small}
  \caption{
  Context-sensitive permission verification and permission inheritance. A dependency without defined permissions is authorized if its caller is authorized. An operation is authorized if all dependencies on the call stack are authorized.
  }
  \label{listing:check-perm-algo}
  \begin{tcolorbox} [width=\linewidth, colback=white!30!white, top=1pt, 
  bottom=1pt, left=2pt, right=2pt]
\begin{minted}[
    fontsize=\scriptsize,
    linenos,
    gobble=2, % Remove unnecessary indentation -- the line numbers make this clear enough
    xleftmargin=0.5cm, % Otherwise we start in the left margin...
    escapeinside=||
]{python}

# In: resource type, specific item, and operation
# Out: True if access is approved, else False
def authorizeAccess(resType, resItem, resOp):
  callStackClasses = getClassesFromCallStack()
  depsPolicies = getPoliciesForDeps(callStackClasses)

  ## This part handles thread-based delegation
  if resType == "thread": # Child thread is created
     # Propagate parent's policy to child thread
     saveThreadParentPolicies(resItem, depsPolicies)
     return True

  ## For access from child threads, 
  ## we check its contextPolicy and that of its parent
  parentContextPolicy = getParentPoliciesForThread()
  if parentContextPolicy != null:
    depsPolicies.extend(parentContextPolicy)

  ## Having obtained the appropriate policy, check it
  for policy in depsPolicies:
    if policy.checkAccess(resType, resItem, resOp):
       continue

    #Policy does not grant access
    if fatal_enforcement_mode:
      raiseException()
    else
      sendAlert()

  
  \end{minted}
\end{tcolorbox}
\end{listing}

\subsection{ZTD-P$_2$: Discovery of Least Privileges}
\label{subsec:discovery-least-privileges}

The huge number of dependencies in modern applications makes it infeasible to manually specify the least privilege policies for an application's dependencies.
\tooldesign infers the dependencies' least privileges from their observed execution behavior.
This approach is motivated by findings from \cref{sec:understanding-sscv} that vulnerabilities commonly expose resources that are not needed by the vulnerable dependencies during legitimate executions. 
\cref{listing:least-priv-discovery-algo} shows the algorithm employed by \tooldesign for discovering and generating the least privilege policies for an application's dependencies. \tooldesign maintains a policy object for each dependency in the application. At runtime, it intercepts invocations of sensitive APIs that can access an operating system resource. It retrieves the classes and dependencies in the call stack and updates the policy of each dependency with the necessary permission and the name of the specific resource object that is being accessed. At intervals, it writes the dependencies' policies to a policy specification file.

We also considered inferring a dependency's least privileges by analyzing the library's source code and identifying invocations of function calls that access different resources. A similar approach was applied by the dependency-level security defenses~\cite{ferreira_containing_2021}. However, this approach does not enable fine-grained policy generation for dependencies as it is difficult to statically determine the specific resource objects that a dependency may dynamically access.
Furthermore, the use of indirection in many languages (\eg C/C++ function pointers, Java reflection) will lead to an imprecise identification of the sensitive APIs that a dependency can access.



\begin{listing}
  \centering
  \captionsetup{font=small}
  \caption{
  Pseudocode for discovering and generating the least privilege policy from the observed execution behavior. 
  }
  \label{listing:least-priv-discovery-algo}
  \begin{tcolorbox} [width=\linewidth, colback=white!30!white, top=1pt, 
  bottom=1pt, left=2pt, right=2pt]
\begin{minted}[
    fontsize=\scriptsize,
    linenos,
    gobble=2, % Remove unnecessary indentation -- the line numbers make this clear enough
    xleftmargin=0.5cm, % Otherwise we start in the left margin...
    escapeinside=||
]{python}


# Function called when a resource access is requested
# In: resource type, specific item, and operation
def onPermissionRequested(resType, resOp, resItem):

  # We get the classes in the call stack
  classes = getClassesFromCallStack()
  
  # For each class, we get its parent dependency
  depPolicies = getPoliciesForDeps(classes)
  # The first dependency is the direct caller. 
  depPolicies[0].addCoarsePerm(resType, resOp)
  depPolicies[0].addFinePerm(resType, resOp, resItem)
  
  # We assign transitive permissions to other deps
  for (i = 1; i < depPolicies.length(); i++) {
    depPolicies[i].addFinePerm(resType, resOp, resItem)
  }

  \end{minted}
\end{tcolorbox}
\end{listing}

\subsection{ZTD-P$_3$: Risk Awareness through Continuous Monitoring}
\label{subsec:continous-monitoring}

As noted in~\cref{subsec:design-reqs}, ZTD requires engineers to continuously monitor their dependencies' resource access.
\tooldesign supports continuous monitoring using the approaches for least privilege discovery and policy enforcement given above.

When run throughout the application's lifetime, the least privilege discovery framework continuously informs software engineers of all resources their dependencies accessed. This improves engineers' risk awareness as they can audit any dependency that accessed a sensitive resource to ensure malicious actors cannot hijack such access. 

In addition, when non-fatal enforcement mode is enabled, ZTD informs software engineers of any unexpected resource access by a dependency, facilitating an investigation. Further investigations that show the access was benign and necessary can lead to improvements of the provided policy.

\section{\toolname: A \tooldesign Impl. for Java Applications}
\label{sec:ztd-java-implementation}

\begin{figure}[h]
 \centering
\captionsetup{font=small}
 \includegraphics[width=0.95\columnwidth]{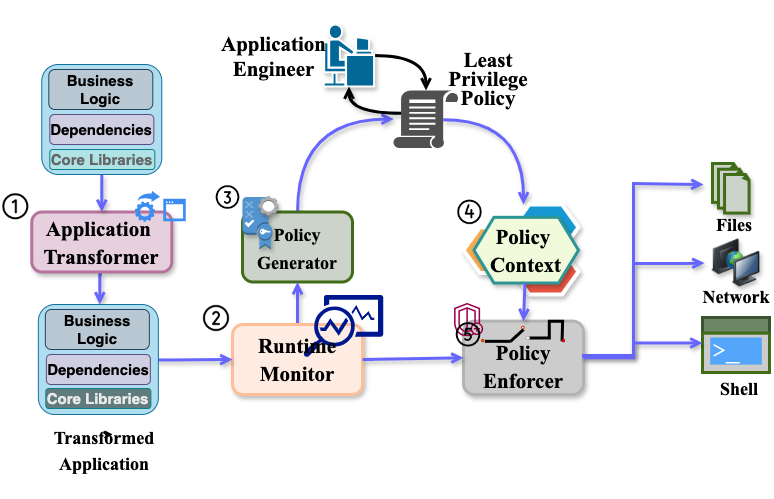}
 \captionof{figure}{
   The \tooldesign design has five components.
   The application transformer instruments the application.
   The runtime monitor tracks dependencies' access to resources.
   The policy generator generates the least privilege policies for dependencies.
   The policy context loads the generated policies.
   The policy enforcer authorizes access. 
   }
 \label{fig:next-jsm-overview}
\end{figure}

\begin{table}
    \centering
    \captionsetup{font=small}
    \caption{
    Java classes and instrumented methods that provide operating system resource access.
    }
    \begin{tabular}{p{2cm}p{2.5cm}p{2cm}}
    \toprule
         \textbf{Operation}&  \textbf{Class}& \textbf{Instrumented Method}\\
    \toprule
         File Read&  FileInputStream& $<$Constructor$>$\\
         File Write&  FileOutputStream& $<$Constructor$>$\\
         Network Connect&  Socket& connect()\\
 Runtime Execution& ProcessBuilder&start()\\
 \bottomrule
    \end{tabular}
    \label{tab:instrumented-methods}
\end{table}

We implemented a prototype of \tooldesign for Java applications: \toolname.
Java applications comprise classes containing the application's business logic, classes from dependencies, and core classes provided by the Java Development Kit (JDK).
The JDK core classes allow the application and its dependencies to access operating system resources and use other features provided by the language. 
\cref{tab:instrumented-methods} shows the core classes for accessing the file, network, and shell system.

\myparagraph{Mapping Dependency Policies to Java Classes at Runtime}
In Java applications, the classes in the call stack do not indicate their parent dependency. However, \tooldesign requires that the policies specified for dependencies apply to all classes within the dependency. Hence, we need to map the classes in the call stack to the specified dependency policy.
\toolname uses the heuristic that Java class names are formed from the directory tree structure containing the class. Classes in the same dependency share a common root directory path and their names share a \textit{common prefix} representing their shared directory path.
We refer to this common prefix as the dependency's \textit{namespace}, as it contains all classes in the dependency.
For dependencies from the \mavencentral registry, this common prefix is usually obtained by combining the unique Group ID and Artifact ID of the library in the registry. Hence, the \code{com.app.bar} policy in \cref{listing:perm-specification} will apply to all classes with names beginning with \code{com.app.bar}.

\subsection{Implementation Details of \toolname's Components}
\toolname is implemented in \totalLoc lines of Java code. As shown in~\cref{fig:next-jsm-overview}, \toolname comprises five components. We discuss the implementation of each component below.

\subsubsection{Application Transformer (AT) \dingone} 
The Application transformer modifies the bytecodes of an application at runtime. 
It is supplied as a command line argument for the \textit{java} command used to execute the application.
It takes in a list of classes and methods to modify.
It uses the Java Instrumentation API~\cite{noauthor_javalanginstrument_nodate_2} to intercept the specified target classes as they are loaded into the Java Virtual Machine and uses the ASM bytecode modification library~\cite{noauthor_asm_nodate} to insert a direct call to the runtime monitor at the start of the specified method. 
By default, the AT instruments the methods in \cref{tab:instrumented-methods}, but it can be configured to transform fewer or more methods, depending on the application's security needs.

\subsubsection{Runtime Monitor \dingtwo}
The runtime monitor is called whenever the application or a dependency attempts to access a protected resource using an instrumented method. Depending on \toolname's configuration, the runtime monitor invokes the policy generator and/or the policy enforcer.

\subsubsection{Policy Generator \dingthree }

The policy generator generates a least-privilege policy for each dependency using the algorithm in \cref{listing:least-priv-discovery-algo}. When a resource is accessed, it adds the required permission to the policy objects of the dependencies on the call stack. The policy objects are written to the least-privilege policy file at specified intervals or during application shutdown. In addition, the generated policy file also informs software engineers of the resources their dependencies access and the potential risks they pose.

\subsubsection{Policy Context \dingfour }
The policy context stores the specified policies for each dependency.
It is implemented using a Patricia tree~\cite{patricia_tree}, where the nodes contain the dot-separated components of the dependency name, and the policies are contained in the leaf nodes. 
As classes do not contain unique identifiers of their parent dependencies, the context store is designed to get the policy of a class' parent dependency using only the class name.
Hence, the policy for classes from the \code{com.foo.baz} dependency can be obtained by using the first three dot-separated components of their class name to transverse the \code{com}, \code{foo}, and \code{baz} nodes in the tree.
When the application is started, \toolname creates the policy context from the provided policy file. At runtime, the policy enforcer uses the context to retrieve the policies that should be applied to each class on the call stack.

\subsubsection{Policy Enforcer \dingfive}

The policy enforcer uses the context-based authorization algorithm in \cref{listing:check-perm-algo} to validate and authorize access to OS resources.
First, we get the set of dependencies for the classes on the call stack, retrieve the policies for these dependencies, and verify that all policies have permission to access the resource.

\subsection{Limitations of \toolname}
\label{subsec:guarantees-limitations}

\vspace{0.1cm}
\begin{itemize}
    \item \textit{Asynchronous delegation:}
    \tooldesign's handling of indirect dependency interactions (\cref{subsec:secure-resource-access}) does not cover delegations between independent threads or processes created by different dependencies. This can occur if a dependency sends user-controlled data to a different thread where the data will influence resource access. 
    A taint-tracking-based technique can associate permissions with communications between threads and processes, which will incur high runtime costs~\cite{brant_challenges_2021}. 
    Alternatively, the permission model in ZTD (\cref{subsubsec:components-perm-model}) could be expanded to recognize threads and processes as resources that require explicit access permissions.

    \item \textit{Use of Native Execution:}
    An attacker can use native libraries to bypass ZTD's policy authorization.
    Our application transformer could protect the API that allows the use of native libraries (e.g. \textit{System.loadLibrary()} in Java). 
    Alternative designs execute native libraries in sandboxes~\cite{vasilakis_breakapp_2018, sun_nativeguard_2014} but this introduces significant performance overhead.

    \item \textit{False positives from incomplete generated policies:}
    We generate least privilege policies for each dependency based on their observed execution behaviors, similar to sandbox mining techniques~\cite{jamrozik_mining_2016, wan_mining_2017, canella_automating_2021, sanders_mining_2019}.
    These policies will block any unexplored behaviors.
    To address this, \tooldesign provides a non-fatal enforcement mode that alerts engineers rather than disrupting the application.

    \item \textit{Namespace Pollution:}
    While rare, multiple dependencies can share the same namespace in a Java application.
    A malicious actor can create a dependency whose class names share the same namespace as another dependency to inherit the legitimate dependency's privileges.
    However, enforcing fine-grained policies will limit the attacker to only resources that the legitimate dependency can access.

    
\end{itemize}

\section{\toolname: Evaluation}
\label{sec:evaluation}

This section evaluates \toolname's effectiveness, runtime costs, and policy configuration costs with the following questions.
\begin{itemize} [leftmargin=25pt, rightmargin=10pt]
    \item [\textbf{RQ1}] \textit{Effectiveness} - 
    Does \toolname prevent SSC vulnerability exploits that access operating system resources?
    \item [\textbf{RQ2}] \textit{Performance cost} - What is the performance cost of \toolname on realistic applications?
    \item [\textbf{RQ3}] \textit{Configuration Effort} - How much effort is required to audit policies generated for realistic applications?
\end{itemize}

\subsection{Setup}

\JD{To address concerns from Reviewer A, we need to acknowledge the threat imposed by reuse here (it affects generalizability / ``validating on the training dataset''). I think what we did was reasonable because our defense is principled, and so we just saved a bunch fo engineering time. But this should be argued explicitly. I put in a Limitations subsection for this purpose.}


\subsubsection{Vulnerability Selection}
\label{subsubsec:eval-vuln-selection}
We use the vulnerabilities from \cref{sscv-study-datasets}.
We selected three vulnerabilities from each vulnerability class (\cref{tab:vulnerability-count}), prioritizing vulnerabilities with a publicly available exploit proof of concept (POC).

\subsubsection{Application Selection}
We used the DaCapo benchmark applications from \cref{sscv-study-datasets}.
To obtain least-privilege policies for each application's dependencies, we used \toolname's policy generator and the DaCapo benchmark suite's test cases.
These tests ran under \toolname without issue.

\subsubsection{Baseline Selection} 
We use the Java Security Manager (JSM)~\cite{noauthor_security_nodate_2} as a performance evaluation baseline as it shares similar requirements with \toolname: authenticating access to OS resources for Java applications.
We could not quantitatively compare to extant dependency-level security defenses (\cref{tab:related-works-sandboxes}), as none target the Java ecosystem. 

\subsection{Methodology}

\noindent

\subsubsection{RQ1: Defense against exploits leveraging OS Resources}
\label{subsubsec:eval-rq1-methodology}

To evaluate RQ1, we took two steps. First, we tested \toolname's ability to mitigate SSC vulnerabilities. For each vulnerability, we created a sample application that depended on the vulnerable library and could be exploited using the available POC. We ran each application with \toolname to check if the exploits were blocked. Our results are in \cref{tab:vuln_exploits}.

Secondly, we assessed whether the least-privilege policy generated by \toolname could prevent exploits in real applications. We injected 9 vulnerabilities into 4 applications: Biojava, Fop, Graphchi, and Zxing. We selected applications built with \maven so that we could easily add vulnerable libraries as dependencies.
We could not exploit 2 vulnerabilities within BioJava because the exploits required Java 8 and Biojava required a minimum of Java 11 to run. 
We used \toolname and the provided workload to generate least-privilege policies for each application and evaluated the effectiveness of the generated policies in blocking exploits of the injected vulnerabilities.
While we injected only nine vulnerabilities in this step, we expect the generated policies to also prevent exploits of remaining vulnerabilities as the vulnerabilities access resources not allowed by the policies. 
In addition, unlike previous security defenses~\cite{ferreira_containing_2021, vasilakis_preventing_2021} that are only evaluated on vulnerabilities recreated in simple applications, we demonstrated \toolname's ability to generate and enforce policies for real applications.

We note that our evaluation uses only vulnerabilities analyzed in \cref{sec:understanding-sscv} and, therefore, the result may not generalize to new vulnerabilities. However, the design of \toolname is based on the general ZTA framework. This should help mitigate all vulnerabilities covered by the threat model in \cref{subsec:threat-model}.

\subsubsection{RQ2: \toolname's Performance Impact}
First, we measured the cost of each instrumented operation with and without \toolname's instrumentation.
We used the Java Microbenchmarking Harness (JMH) library~\cite{noauthor_openjdk_nodate_2} maintained by OpenJDK to avoid microbenchmarking pitfalls~\cite{costa_whats_2021}. 

Second, we measured performance while increasing the number of dependencies (\textit{expected: constant}) and the number of classes from different dependencies in the call stack during a method's invocation (\textit{expected: linear}).

Third, we profiled the applications.
We measured the execution time of each application without sandboxing, with \toolname (configured in four modes), and with JSM.
We used the \textit{converge} feature of the DaCapo benchmark harness~\footnote{An application is executed until three executions have times within 3\%, then the next iteration's time is kept. We repeated the average of 3 trials.}.


\subsubsection{RQ3: \toolname's Policy Configuration Effort}

We measured the effort to audit the least-privilege policies generated by \toolname for DaCapo benchmark applications.
As default policies are automatically generated, engineers should not need to master a new policy language.
Therefore, the configuration effort will be spent mainly to audit and refine the generated policies.

\JD{I added a paragraph break here. I think this improves readability. But if it breaks the page limit, please remove the paragraph break!}
To the best of our knowledge, there is no existing metric for configuration effort in this context. We therefore estimate that the configuration effort depends on two factors:
  (1) the number of dependencies for which \toolname generated a non-empty policy,
  and
  (2) the number of permissions provided in each such policy.
The first metric captures the number of policies that engineers would need to study.
The engineer may also need to assess if each provided permission is valid, so the second metric captures the level of analysis that they must perform for each dependency.
For each application, we report the number of dependency policies generated and the average number of permissions per policy.


\subsection{Results}

\subsubsection{RQ1: Defense against exploits leveraging OS Resources}
\label{subsubsec:rq1}

\begin{table}[t]
    \centering
    \captionsetup{font=small}
    \caption{
    \toolname's performance in preventing host-system-compromising exploits.
    It successfully blocks selected exploits. We injected 7 vulns. in 4 applications and 2 vulnerabilities in only 3 applications. Their exploits were blocked.
    A `--' means we did not inject this vulnerability.
    }
    \label{tab:vuln_exploits}
    \begin{tabular}{cccccc}
    \toprule
         \textbf{CVE ID}& \textbf{Vuln. class} & \textbf{Impact} & \textbf{S. Apps} & \textbf{D. Apps} \\
         
     \midrule

     CVE-2020-9547 & Des. & Code Ex.  & \cmark & - \\
     CVE-2020-8441 & Des. & Code Ex.  & \cmark & 3/3 \\
     CVE-2022-36944 & Des. & File Man. & \cmark & - \\

     CVE-2021-44228 & Code Inj. & Code Ex.  & \cmark & 3/3\\
     CVE-2022-33980 & Code Inj. & Code Ex.  & \cmark 
 & 4/4\\
     CVE-2022-22963 & Code Inj. & Code Ex. & \cmark & - \\

     CVE-2023-39021 & Comm. Inj. & Code Ex.  & \cmark & 4/4 \\
     CVE-2023-39020 & Comm. Inj. & Code Ex.  & \cmark & 4/4 \\
     CVE-2022-25914 & Comm. Inj. & Code Ex. & \cmark & 4/4 \\

     CVE-2022-0839 & XXE & Data Exf. &  \cmark & 4/4 \\
     CVE-2021-23463 & XXE & Data Exf. &  \cmark & - \\
     CVE-2019-10172 & XXE & Data Exf. & \cmark & 4/4 \\

     CVE-2022-4244 & Path Trav. & File Man. &  \cmark & 4/4 \\
     CVE-2020-17518 & Path Trav. & File Man. & \cmark & - \\
     CVE-2020-17519 & Path Trav. & File Exf. & \cmark & - \\
         
         \bottomrule    
    \end{tabular}
\end{table}

\cref{tab:vuln_exploits} shows details of the vulnerabilities we recreated. As shown, \toolname blocked the exploits of vulnerabilities in the sample applications, When applied to real applications, we find that the least privilege policies of all 4 applications were sufficient to prevent the 9 vulnerability exploits.


\subsubsection{RQ2: \toolname Performance Impact}
\label{subsec:perf-eval}

\begin{table}
    \centering
    \captionsetup{font=small}
    \caption{
    Microbenchmark of security-sensitive operations.
    Uncertainty is reported after the measurements converge.
    }
    \label{tab:microbenchmark_results}
    \begin{tabular}{p{1.8cm}p{2cm}p{3.5cm}}
    \toprule
         \textbf{Operation}&  \textbf{Vanilla ($\mu$s)}&  \textbf{\toolname ($\mu$s)}\\
     \midrule
         File Read&  $7.96 \pm 0.27$ &  $23.62 \pm 0.20$ (196.7\%)\\
         File Write&  $33.78 \pm 1.73$&  $33.23 \pm 0.56$ (-1.6\%)\\
         Socket Connect&  $71.23 \pm 2.56$& $72.19 \pm 2.34$ (1.3\%) \\
         Shell Execution&  $320.69 \pm 27.9$ & $345.70 \pm 3.98$ (7.8\%) \\
         \bottomrule
    \end{tabular}
    
\end{table}

\begin{figure}[h]
 \centering
 \captionsetup{font=small}
 \includegraphics[width=0.75\linewidth]{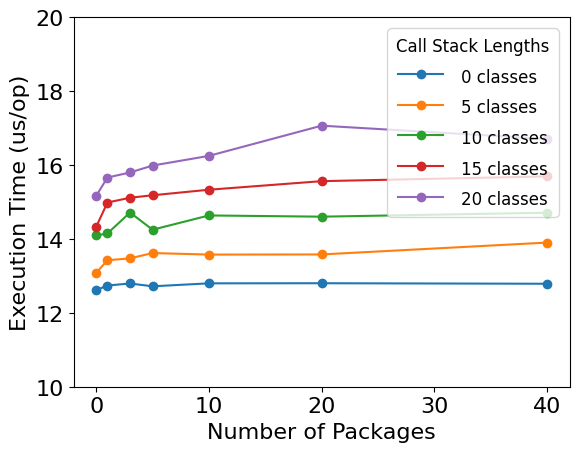}
 \captionof{figure}{
   Microbenchmarking results for the policy authorization operation, varying the number of dependencies in the application and the call stack sizes.
   As predicted, the execution time is constant with the dependency count and linear to the call stack lengths.
   }
 \label{fig:check-perm-overhead}
\end{figure}

\begin{table}[t]
    \centering
    \captionsetup{font=small}
    \caption{
    Execution time overheads and \toolname configuration effort on the DaCapo applications. The table shows dependency and authorization call counts, the overhead of \toolname and JSM, and the configuration effort for each app (x/y means x dependencies need policies and each policy provides an average of y permissions).
    \mbox{`--': data} unavailable because (column 2) application is not built with \maven/Gradle, or (column 6) JSM could not execute.
    }
    \label{tab:profiling_results}
    \begin{tabular}{p{1cm}p{0.5cm}p{0.6cm}|C{0.7cm}C{0.9cm}C{0.85cm}|C{1.0cm}}
    \toprule
         \textbf{App Name}&  \textbf{Deps count} & \textbf{Auth calls} & \textbf{Vanilla} & \textbf{\toolname} &  \textbf{JSM} & \textbf{Config Effort}\\
         \midrule
         Avrora & 3 & 6 &  10,036  & 0.52\% & 0.18\% & 2/2 \\ 
         Batik & 24 & 16 & 1,605 & 0.83\% & -- & 3/2.7 \\ 
         Biojava & 34 & 6 & 10,589 & -0.56\% & -0.75\% & 7/2 \\ 
         Eclipse & -- & 23k & 16,464 & 2.91\% & 201.04\%  & 9/3 \\ 
         Fop & 30 & 94 & 734 & 0.23\% & 28.60\% & 8/2.5 \\ 
         Graphchi & 40 & 97 & 5,245 & -3.18\% & -0.19\% & 3/2 \\ 
         H2 & -- & 10 & 3,599  & 0.60\% & -0.27\% & 2/2 \\
         Luindex & -- & 11 & 5,933  & -0.36\% & 14.22\% & 4/2 \\
         Lusearch & -- & 6k & 3,534 & -0.48\% & -0.50\% & 4/2 \\
         PMD & 47 & 12 & 1,953 & -1.02\% & 0.96\% & 3/2 \\ 
         Spring & 111 & -- & 5,800 & -1.17\% & -- & 17/1.6 \\ 
         Tomcat & -- & 1k & 4,229 & 0.50\% & 10.99\% & 19/2.3 \\
         Tradebeans & -- & 42 & 23,401 & 0.07\% & -- & 13/2.9 \\
         Tradesoap & -- & 37 & 13,129 & 0.10\% & -- & 13/3 \\
         Xalan & -- & 9k & 978 & -2.04\% & -0.58\% & 5/2 \\
         Zxing & 5 & 2.5k & 1,290 & 2.53\% & -0.96\% & 3/1.7 \\ 
         \midrule
         \textbf{Medians} & -- &  -- & -- & -0.03\% & 21.06\% & 5/2 \\
         \bottomrule    
    \end{tabular}
\end{table}

\cref{tab:microbenchmark_results} shows the results of microbenchmarks for 4 operations: file read, file write, socket connection, and shell execution. 
\toolname introduces modest overhead in file write, socket, and shell operations.
For scaling,~\cref{fig:check-perm-overhead} shows that \toolname's performance does not depend on the number of dependencies and is marginally affected by call stack depth. 

\cref{tab:profiling_results} shows the profile results on applications from the DaCapo benchmark suite.
\toolname introduces no noticeable overhead on 7 applications,
with $<$1\% overhead on 13 of the 16 applications.
The JSM, despite operating at a higher abstraction level, introduces $>$10\% overhead on four of sixteen applications, with $\sim$200\% overhead in Eclipse.

We attribute \toolname's minimal overhead, compared to the JSM, to two factors.
First, \toolname performs policy authorization checks at the dependency granularity, \ie once per dependency in the call stack (\cref{listing:check-perm-algo}).
In contrast, the JSM checks permissions at the \textit{class} granularity in the call stack. 
Second, \toolname performs less frequent policy authorization checks.
The default implementation enforces only 4 permissions while the JSM enforces 28 permissions and protects 100+ methods~\cite{oracle2014permissions}.


\subsubsection{RQ3: \toolname's Configuration Effort}
The last column of \cref{tab:profiling_results} shows the number of dependency policies generated for each application and the average number of permissions provided in each policy. 9 of the 16 applications have less than 5 dependencies that require policy specification. Policies provide only 1-3 permissions to the dependency. All policies were generated by \toolname without manual effort and executed on applications without failure.

Note that the count of policies and permissions in \cref{tab:profiling_results} includes both direct and transitive permissions. Hence, compared to the data in \cref{tab:profiling_results}, only fewer dependencies directly accessed any operating system resources.

\section{Discussion}
This section discusses \toolname's application to software applications (\cref{subsec:ztd-real-apps}) and other future directions (\cref{subsec:ztd-future-directions}).

\subsection{Applying \toolname to Software Applications}
\label{subsec:ztd-real-apps}

\subsubsection{Potential Usage Scenarios of \toolname}
ZTD aims to protect software applications from SSC vulnerabilities. We foresee three potential usage scenarios for \tooldesign based on the risk tolerance and reliability needs of an application.

\myparagraph{Proactive Security}
Applications run in fatal enforcement mode (\cref{subsec:enforcing-ztd}). The engineering team observes dependency resource access, generates policies, and enforces them to block abnormal resource access. This mode provides high security, but enforcing incomplete policies can lead to false positives and disrupt the application. This mode would have prevented the Equifax breach (CVE-2023-50164)~\cite{noauthor_equifax_nodate_2} and attacks due to the Log4J vulnerability (CVE-2021-44228)~\cite{Log4ShellSCMag2022}.

\myparagraph{Reactive Security}
Applications generate policies and enforce them nonfatally (\cref{subsec:enforcing-ztd}). 
This alerts engineers to unauthorized access without disrupting the application.
This mode would have detected the Equifax breach earlier, preventing the reported 76-day dwell time~\cite{equifax_breach} and limited its impact.

\myparagraph{Emergency Security}
Applications have policy discovery enabled (\cref{subsec:discovery-least-privileges}) but no enforcement. When new vulnerabilities are reported in a dependency, the policies discovered for that dependency can be modified and enforced to mitigate the vulnerability by blocking access to any exposed resource. This mode can act as \textit{emergency first-aid} and would have prevented software engineers from shutting down applications after the discovery of CVE-2021-44228~\cite{log4j_shutdown_sas_2021}.

\subsubsection{Test cases for generating least-privilege policies} 
In the usage scenarios described, the engineering team responsible for the application is tasked with selecting the test cases for policy generation. They can use available workloads for end-to-end tests or use fuzzing to generate new high-coverage test cases~\cite{olsthoorn_syntest-solidity_2022, liu_pmfuzz_2021}.
Alternatively, they can run their application in \tooldesign's policy discovery mode until they feel confident that the necessary functionalities have been exercised. 
This process is performed once for the entire application and needs only to be repeated when new functionalities or dependencies are introduced. 
Furthermore, discovered least-privilege policies are unlikely to be affected by dependency version updates, as such updates rarely introduce new permission requirements~\cite{ferreira_containing_2021}.
In the future, we intend to implement lightweight coverage metrics that help engineers decide when sufficient behaviors have been covered.

\subsubsection{Barriers to \toolname Adoption}
While \toolname addresses SSC vulnerabilities, we foresee two barriers to adoption.

\begin{enumerate}
    \renewcommand{\labelenumi}{\alph{enumi})}   
    \item Engineers may fear that runtime modifications might introduce errors~\cite{liu_instruguard_2021}. \toolname mitigates this concern by modifying only a few (4) classes.
    All are in JDK core.
    \item False positives may occur if \toolname encounters legitimate resource accesses that were not previously observed.
    We mitigate this by adding the non-fatal enforcement mode where any false positives will not disrupt the application.
    A possible \toolname utility would be a code coverage metric to assess if sufficient program behavior has been covered.
\end{enumerate}

\subsubsection{\toolname vs. the Java Security Manager (JSM)}
\label{subsubsec:jsm-comparison}
Both \toolname and the (deprecated) JSM address Java application vulnerabilities via resource access authorization.
As noted in~\cref{sec:RelatedWork}, JSM had high runtime costs (up to 200\%) as well as high policy complexity: JSM protected 100+ operations with 28 different permission types~\cite{oracle2014permissions}.
Based on common SSC vulnerabilities, \toolname protects 3 sensitive resources by default, allowing lower performance costs and policy complexity.
However, \toolname's application transformer (\cref{sec:ztd-java-implementation}) can also be configured to protect a different set of resources according to the security and performance needs of the application.
Furthermore, \toolname overcomes the usability flaws that led to the lack of use of JSM and subsequent deprecation~\cite{sean_mullan_jep_nodate_2}. It features a flexible permission model (\cref{subsec:ztd-policy-auth}) that allows coarse-grained policies for easy-to-develop policies and fine-grained policies for stronger security, easier programming with automated policy discovery (\cref{subsec:discovery-least-privileges}), and low performance cost (\cref{subsec:perf-eval}). 
While \toolname is designed to address vulnerabilities in an application's dependencies, it can also be extended to mitigate the application's own vulnerabilities, serving as an effective replacement for the deprecated JSM.

\subsubsection{Design Choice: Protect Resources or Operations}

In line with ZTA, \tooldesign protects functions that access resources. This approach differs from JSM~\cite{noauthor_security_nodate_2} and some other dependency-level defenses~\cite{vasilakis_preventing_2021, vasilakis_breakapp_2018}, which protect functions performing sensitive operations.
Protecting sensitive operations provides finer control over dependencies' behavior. However, the size, complexity, and diversity of applications and programming languages mean there are many `sensitive' operations, each requiring a different permission. As seen with JSM (\cref{subsubsec:jsm-comparison}), this leads to increased policy complexity and runtime performance overhead.
To provide a balance, \tooldesign offers a customizable application transformer for adding or disabling permissions.

\subsection{Future Directions}
\label{subsec:ztd-future-directions}

\subsubsection{Improving the visibility of third-party library capabilities}

Recent supply chain vulnerabilities like Log4J/Log4Shell~\cite{Log4ShellSCMag2022} highlight the need for third-party library engineers to communicate the functionalities and resource usage of third-party libraries.
This practice would help application engineers assess their risks, and discourage the provision of unnecessary OS resource capabilities to these libraries.
Initiatives like Software Bill of Materials~\cite{noauthor_software_nodate_2} and Open-source Insights~\cite{noauthor_open_nodate_2} could make this task machine-checkable if these approaches were extended to document the capabilities of third-party libraries.
\toolname's policy generation component can assist engineers in documenting their libraries' capabilities and permissions.

\subsubsection{Application to Other Ecosystems and Programming Langs}


Supply chain attacks in the Java ecosystem exploited accidental vulnerabilities~\cite{noauthor_equifax_nodate, CISALog4jReport2022}, unlike attacks in the JavaScript and Python ecosystems which are often caused by malicious vulnerabilities~\cite{ohm_backstabbers_2020}. This emphasizes the need for runtime defenses that protect apps from vulnerable dependencies in addition to preventing the use of malicious libraries.

The \ztd concept is language-agnostic and protects applications relying on third-party libraries (\cref{sec:design-implementation}).
\toolname instrument functions in a language's core libraries that interact with the operating system's resources (\cref{sec:design-implementation}).
With minimal adjustments, the tool can be used with other JVM-based languages such as Kotlin~\cite{kotlin} and Groovy~\cite{apache_groovy}.
Application to other languages is future work.

\section{Conclusion}

Software supply chain vulnerabilities and attacks are increasing.
Using NIST's \zta as a framework, we show that existing security defenses are incomplete.
We propose, design, implement, and evaluate the \textit{\ztd} approach.
\ztd
  places zero implicit trust in third-party software libraries,
  and
  can mitigate risks and protect operating system resources without impeding the application's normal behavior.
We design a low-cost, low-effort system, \tooldesign, enabling \ztd's application to software.
Our prototype for Java applications, \toolname, effectively mitigates supply chain vulnerabilities with minimal runtime cost and ease of configuration.
In summary, \ztd provides a robust defense against risks from vulnerable dependencies.

\section*{Data Availability}

Our artifact is at: https://doi.org/10.5281/zenodo.14436182.
It has
vulnerability data,
\toolname source, and
experiments.


\clearpage

{\footnotesize \bibliographystyle{unsrt}
\bibliography{extra-refs, master, GrayDuality}}

\end{document}